# Nudging Consent & the New Opt-Out System to the Processing of Health Data in England

**Janos Meszaros, Chih-hsing Ho and Marcelo Corrales Compagnucci**

**Abstract:** This chapter examines the challenges of the revised opt-out system and the secondary use of health data in England. The analysis of this data could be very valuable for science and medical treatment as well as for the discovery of new drugs. For this reason, the UK government established the "care.data program" in 2013. The aim of the project was to build a central nationwide database for research and policy planning. However, the processing of personal data was planned without proper public engagement. Research has suggested that IT companies – such as in the Google DeepMind deal case – had access to other kinds of sensitive data and failed to comply with data protection law. Since May 2018, the government has launched the "national data opt-out" (ND opt-out) system with the hope of regaining public trust. Nevertheless, there are no evidence of significant changes in the ND opt-out, compared to the previous opt-out system. Neither in the use of secondary data, nor in the choices that patients can make. The only notorious difference seems to be in the way that these options are communicated and framed to the patients. Most importantly, according to the new ND opt-out, the type-1 opt-out option – which is the only choice that truly stops data from being shared outside direct care – will be removed in 2020. According to the Behavioral Law and Economics literature (Nudge Theory), default rules – such as the revised opt-out system in England – are very powerful, because people tend to stick to the default choices made readily available to them. The crucial question analyzed in this chapter is whether it is desirable for the UK government to stop promoting the type-1 opt-outs, and whether this could be seen as a kind of "hard paternalism."

# 1 Introduction

Governments are always actively seeking to enable efficient healthcare systems with the aim of improving the quality of care while reducing public spending.[1] The subject of this chapter is the secondary use of health data in England, which is one way of reaching these goals. The secondary use of health data refers to the processing of data collected during direct care for new purposes, such as research and policy planning.[2] England has adopted a new opt-out system called "national data opt-out" (ND opt-out) and it is available since May 2018.[3]

Default rules – such as the ND opt-out system in England – are very powerful. The reason is that people tend to stick to the default option and choosing a different option requires an active decision and further deliberation costs. In other words, people tend to prefer the easiest option. In this case, the option which does not require mental effort such as in the opt-out systems. However, the further use of health data poses complex ethical,[4] legal and technical challenges.[5]

Default rules can create a lot of good, but also do a lot of harm. This is one of the key conceptual arguments of the Behavioral Law and Economics literature that blend insights from cognitive psychology and economics.[6] It takes into account the psychological traits of human behavior and a variety of other factors such as emotional, social and cognitive as the overarching framework to discuss legal issues.[7] The subject of this chapter is about architectures, freedom of choice, and the legitimate ways of the UK government to nudge its citizens – as a new form of "hard paternalism."

The chapter is divided into 7 sections. After this introduction, Sect. 2 explains the main tenets of Behavioral Law and Economics. It uses real life examples to illustrate the pervasive nature of nudges – in particular, default rules – and choice architectures which are everywhere, influencing inadvertently the decisions that people make. Section 3, discusses nudging techniques in the healthcare sector, in particular opt-out systems. Section 4, explains the National Health Service (NHS) and the opt-out system in England. This section is divided into two main parts. The first part revisits the old opt-out system and explains the new types of opt-out rules set out in the new ND-opt-out. The second part, focuses on the way this information is presented and framed to the patients. Section 5, provides some statistics on opt-outs based on recent data released by the NHS.

---

[1] Deloitte (2016), p. 3.
[2] Institute of Medicine (2013); Hanney and González-Block (2015), pp. 1-4.
[3] See Digital NHS UK. Available at: https://digital.nhs.uk/national-data-opt-out. Accessed 10 June 2019.
[4] Institute of Medicine (2013); Sunil, Douglas and Leo (2016), pp. 17-25; Safran et al. (2007), pp. 1-9.
[5] Safran et al. (2007), pp. 1-9.
[6] Given the importance of this new field of law, the Government of the United Kingdom Cabinet Office established the "Behavioral Insight Team" (BIT) – unofficially known as the "Nudge Unit." The BIT was originally a governmental organization set up to apply insights from behavioral economics to improve public policy and services.[6] Recently, the BIT became a limited company and it is co-owned by the government. Since the BIT was spun off as a social purpose company, it has given birth to a global movement that now spans 153 countries. See, The Behavioral Insight Team. Available at: https://www.bi.team. Accessed 10 June 2019.
[7] Angner and Loewenstein (2016) pp. 1-56; see generally, Zeiler and Teitelbaum (2015); Minton and Kahle (2013).

Section 6, delves into details concerning the theoretical discourse of the so-called "Libertarian Paternalism" in Cass Sunstein's narrative. According to Sunstein, there are different kinds of paternalisms. The main distinction relevant to the discussion of this chapter is the difference between hard vs. soft paternalism. The first one coerces individual freedom. Hard paternalisms are, therefore, not desirable. The latter provides freedom of choice. This is the kind of paternalism advocated in the last section of this chapter (Sect. 7), which concludes with the opinion that the type-1 opt-out should not be ruled-out from the revised opt-out system in England.

## 2 Behavioral Law and Economics, Choice Architectures & Default Rules as Prime Nudges

Behavioral Law and Economics became very popular and entered the mainstream of modern law and economics thanks to the works of Richard Thaler and Cass Sunstein. The first won the Nobel Prize in Economics with Nudge Theory in 2017. The main postulate of this theory is that improved choices and information disclosure could softly *nudge* (push or poke gently)[8] individuals to improve decision-making and welfare.[9]

Real-world illustrations of nudges can be found everywhere in our daily lives. Urinals at Amsterdam Schiphol airport with images of a fly just above the drain are one of such examples. According to Nudge Theory, the image of a fly would attract the attention of men and prompt them with a target at which they would aim. This experiment showed "spillage" on the bathroom floor was considerably reduced by 50 to 80 %.[10]

Another typical example of a *nudge* in our daily life is a cafeteria. Think of the manager of the cafeteria who has the freedom to arrange the food in certain places. She could place the food in a place that is more visible to people in order to affect their decisions. Putting the salad at the entrance and in a visible place, would increase the likelihood that customers would choose the salad first as a healthier option.[11] To count as a nudge, "the intervention must be easy and cheap to avoid."[12] Placing the salad at eye level is a nudge. Banning junk food however is a mandate.[13]

Empirical studies were carried out at two college campuses – the University of Connecticut and Alfred University in New York – by students and managers interested in seeing how re-arranging cafeteria options would influence student behavior. Their interest was not healthy eating as in the previous cafeteria example, but waste. They realized that it was very easy to load trays with food that ended up as a waste. Therefore, they ran a "trayless" experiment for a couple of days and noticed that food and beverage waste dropped between 30 and 50 %. This amounts to 2 tons of food and about 424 liters of liquid waste saved on a weekly basis.[14]

---

[8] See English Collins Dictionary (Nudge).
[9] Corrales and Kousiouris (2017), p. 161.
[10] Corrales and Jurčys (2016), p. 533.
[11] Corrales and Jurčys (2016), p. 533.
[12] *Thaler and Sunstein* (2009), p. 6.
[13] Coggon, Syrett and Vienns (2017), p. 177.

These examples take us directly to the definition of a "choice architect." A choice architect, is any person who changes "every small feature in the environment that attracts our attention and influences the decision that we make."[15] For Thaler and Sunstein, a nudge is "any aspect of the choice architecture that alters people's behavior in a predictable way without forbidding any options or significantly changing their economic incentives."[16] In other words, Nudge Theory is mainly about *designing* choices that influence and prompt individuals to improve the decisions that they make.[17]

A canonical example of a nudge is the Global Positioning System (GPS) which helps individuals find the best and shortest route. An interesting feature of the GPS is that individuals can always take another route and the GPS can easily track the location again and re-direct them with the route using signals from satellites. In this respect, the GPS system does not coerce individuals to take one particular route. It is up to the users to follow the directions of the GPS or not. This is one of the main characteristics of a nudge. A nudge never overrides individual freedom.[18]

Nudging is nothing new. The private sector has been nudging consumers for decades. Marketing agencies have always used different nudging techniques to attract the attention of their customers and influence their behavior in order to sell their products. Nudges can be very helpful for individuals and society. Some of these nudges, however, may be regarded to be more controversial than others. For example, road signs are undeniably helpful for the community. They can be hardly regarded to be controversial. They give instructions to people and warn them to drive more carefully. However, if road signs are put in the wrong place, they could be dangerous and create accidents.[19]

Choice architects have the responsibility of organizing the context in which people make decisions.[20] Thus, choice architectures embrace the idea of nudges. The most powerful nudges are: warning signals, information disclosures and *default rules* (emphasis added). In this chapter, we will focus on default rules as prime nudges. According to the Behavioral Law and Economics literature, they are inevitable and they are everywhere.[21]

A classic example to illustrate a nudge as a default rule is the hardware and software of printer machines. Users can choose between single or double-sided printing. Single-side printing would obviously require more paper and ink, while double-sided printing would significantly reduce the costs expenditures of an organization. Rutgers University ran an experiment in its New Brunswick campus. They simply changed the default settings of all printers from one-sided to double-sided and saved over eighty-nine million sheets of paper during the first years of the conservation program. This amounted to a 44 % reduction which is tantamount to 4,650 trees. This zero-cost option is a good example of how changing a small feature in the design of the computer software and hardware architecture can make a big difference.[22]

---

[14] See Nudging in the Cafeteria (2008). Available at: https://nudges.wordpress.com/2008/04/17/nudging-in-the-cafeteria/. Accessed 10 June 2019.
[15] Willis O (2015) Behavioral Economics For Better Decisions, ABC.net. https://www.abc.net.au/radionational/programs/allinthemind/better-life-decisions-with-behavioural-economics/6798918. Accessed 10 June 2019.
[16] Thaler and Sunstein (2009) p. 6.
[17] Businessballs.com. Nudge theory. Available at: https://www.businessballs.com/improving-workplace-performance/nudge-theory/. Accessed 10 June 2019.
[18] Corrales and Kousiouris (2017), pp. 165-166.
[19] See, e.g., generally, Jamson (2013), p. 298.
[20] Thaler and Sunstein (2009), p. 3.
[21] Sunstein (2014), pp. 1-30, 179.

## 3 Nudges in the Healthcare Sector & Opt-Out Systems

A cursory look at behavioral economic insights in the healthcare sector suggests that these interventions tend to be small and often set as default rules. Subtle changes in how choices are presented, for instance, as opt-in or opt-out can make big changes in the behavior of participants. A very good example to illustrate this are default rules related to the postmortem organ donation.[23] Two main default system exist at the global scale: i) *opt-in system:* which requires explicit consent from the deceased, and; ii) *opt-out system:* whereby consent is automatically assumed.[24] The latter means that the deceased is a donor by default.[25]

The procedure for opt-in and opt-out differs greatly from country to country. In the United States the deceased must have previously signed up in a state registry,[26] whereas in other countries such as in Japan and most European Member States, citizens have the option to check a box as an opt-in or opt-out rule when they have to renew their driving license. Opt-out default systems make the percentage of organ donation much higher than in opt-in systems.[27] For example, countries such as Spain, Austria, France, Hungary, Poland and Portugal, have all implemented opt-out systems and the number of organ donation increased exponentially to 99%[28] in comparison to other countries such as Denmark (4.35%) and the Netherlands (27.5%)[29] which have opt-in systems.[30]

Decision-making using opt-in or opt-out rules could also be attributed to the culture of a society, which may incrementally change its perception based on user experience. Empirical studies conducted in countries such as Germany, United States and Austria, revealed that signing up for an organ donation in an opt-in system was generally considered to be a virtuous act of benevolence, whereas abstaining to donate under an opt-out system was commonly viewed as egotistic and antisocial.[31]

---

[22] Stoknes (2015), p. 25; see also Sunstein (2016).
[23] Ben-Porath (2010), p. 11.
[24] Heshmat (2015), p. 243.
[25] Corrales, Jurčys and Kousiouris (2019), p. 197.
[26] Detels and Gulliford (2015), p. 782.
[27] See John et al. (2013), p. 104; Quigley and Stokes (2015), p. 64; Thaler (2009); Hamilton and Zufiaurre (2014), p. 18.
[28] Leitzel (2015), p. 137.
[29] Shafir (2013) (ed), p. 496.
[30] Corrales, Jurčys and Kousiouris (2019), p. 197.
[31] Zamir (2015), p. 103; see, also, generally, Davidai, Gilovich and Ross (2012), pp. 15201-15205.

In a recent blog by surgeon Ara Darzi – director of the Institute of Global Health Innovation of the Imperial College London – he explains the potential of behavioral economics in the field of public health and how this could be effectively applied in cancer screening. Cancer survival is lower in the UK in comparison to other countries and screening significantly reduces the morbidity and mortality. One approach to raise cancer screening is to link the screening test to other tests that people are more familiar with, such as dental check-ups or annual vehicle testing (MOT test in the UK).[32] This is just another good example of how nudging and default rules can make a beneficial impact in the health care service. Below we explain the role of the National Health Service and the recent developments of the opt-out system in the England.

## 4 The National Health Service & Opt-Out System in England

The National Health Service (NHS) provides universal and free public health services in England. As a result, NHS data provide a valuable resource of routinely collected primary (e.g., visiting general practitioner (GP) practices) and secondary (e.g., hospital admissions, outpatient appointments, accident & emergency attendances) healthcare data covering almost the whole population of England.[33] Three broad categories of data are collected from patients during direct care in England: i) basic personal data such as age and gender; ii) medical information such as diagnosis; and, iii) administrative information (e.g., waiting times).[34]

This information is protected by the common law duty of confidentiality (CLDC). Only in special cases can this information be processed for a new purpose, without consent. These situations might be a court order or overriding public interest (e.g., epidemics). Furthermore, Section 251 of the 2006 National Health Service Act allows the Secretary of State for Health to make regulations that bypass the CLDC for defined medical purposes. The Health Service (Control of Patient Information) Regulations 2002 play this role, as they allow the disclosure of confidential patient information without consent.[35] The main reason for these disclosures is that seeking consent would require disproportionate effort or it would be impossible, and processing anonymous data might be not useful for research and planning in many cases.[36]

Against this legal background, the care.data program was initiated in 2013, aiming at extracting patient information from direct care providers without patient consent, to build a central nationwide database for research and policy planning.[37] However, the processing has been planned without proper public engagement, and the central database was aimed to be used by public and private third-party users, such as IT companies. The care.data program has raised serious public concerns[38] and it has been paused several times.[39] During this period, the National Data Guardian[40] (NDG) started to investigate the care.data program and developed models for consent and opt-out.[41] The NDG published the "Caldicott Review" in 2016, a report that highlighted the issues about the secondary use of health data in the UK, and provided several recommendations for processing, consent and opt-out.[42] After the publication of the report, the NHS in England canceled the care.data program.[43] Despite the failure, the government remained committed to realizing the benefits of the secondary use of health data without consent; thus, the government implemented a new national data opt-out (ND opt-out) system in 2018.

---

[32] Darzi (2017).
[33] Piel et al. (2018), pp. 594-600.
[34] NHS factsheets for health and care staff, Factsheet 1B – Types of data used and legal protection in place (2018), p. 1. Available at: https://digital.nhs.uk/services/national-data-opt-out-programme/guidance-for-health-and-care-staff. Accessed 10 June 2019.
[35] Mészáros and Ho (2019), p. 13.
[36] Lee, Heilig and White (2012), pp. 38-44.
[37] For more information behind the NHS plans for the secondary use of health data, before the creation of care.data program, see Department of Health (2006); see also Department of Health (2015).

### *4.1 The Opt-Out System in England*

The state has the power to bypass patient consent in some circumstances, such as improving healthcare and medical research for the public interest. On the other hand, the citizens' autonomy might be maintained by giving them the choice to opt-out. Thus, the purpose of the opt-out system is to use confidential data without the burden of seeking explicit consent, while respecting peoples' autonomy. There might be cases, however, when opt-outs may seriously hinder data integrity, thus they cannot be applied.[44] The number of opt-outs can be decreased with nudging techniques against this option by, for example, highlighting and framing the importance of the secondary use of health data. This is where Nudge Theory comes to the fore. The UK Government dedicated efforts to convince the patients about the value of research and policy planning. Nevertheless, the public dissatisfaction with the Google DeepMind patient data deal[45] and several other studies[46] indicated that the UK citizens might have concerns about whether their sensitive data is processed for a new purpose outside of direct care.

In the Google DeepMind patient data deal, for instance, Google's artificial intelligence firm was allowed to access health data from over 1.6 million patients to develop an app monitoring kidney disease called "Streams."[47] Public concerns and corroborative research studies suggested that the Google DeepMind deal had access to other kinds of sensitive data and failed to comply with data protection law.[48]

---

[38] Sterckx and Cockbain (2014), pp. 227-228; Mori (2016).
[39] Vezyridis and Timmons (2017), p. 2.
[40] The National Data Guardian (NDG) advises and challenges the health and care system in the UK to help ensure that citizens' confidential information is safeguarded securely and used properly. Available at: https://www.gov.uk/government/organisations/national-data-guardian/about. Accessed 10 June 2019.
[41] National Data Guardian for Health and Care (2016), pp. 1-56.
[42] National Data Guardian for Health and Care (2016), pp. 6-9.
[43] Department of Health and Social Care (2016). Available at: https://www.gov.uk/government/speeches/review-of-health-and-care-data-security-and-consent. Accessed 10 June 2019.
[44] Jones et al. (2017), pp. 43-50.
[45] The ICO has ruled the Royal Free NHS Foundation Trust failed to comply with the Data Protection Act when it provided patient details to Google DeepMind. Available at: https://ico.org.uk/about-the-ico/news-and-events/news-and-blogs/2017/07/royal-free-google-deepmind-trial-failed-to-comply-with-data-protection-law/. Accessed 10 June 2019.
[46] Stockdale, Cassell and Ford (2018), pp. 1-25; Wyatt, Cook and McKevitt (2018), pp. 1-8; Aitken et al. (2016), p. 73.

The NDG report and the failure of the care.data program lead to a change in the opt-out system in England to earn back the public trust. The "old" opt-out system consisted of two choices for the citizens: type-1 and type-2 opt-outs. The "new" system is the ND opt-out, and it is publicly available online since May 2018. However, as we will point out in the conclusion of this chapter, there is neither a significant change in the secondary use of data, nor in the choices that patients can make. The only notorious difference is how these options are communicated to the patients.

### 4.1.1 The "Old" Opt-Out System

In England, there have been two options for citizens to prevent the secondary use of their confidential information. The type 1 opt-out prevented the information being shared outside the GPs practice for purposes other than direct care, while the type 2 opt-out prevented information being shared outside the NHS Digital[49] for purposes beyond the individual's direct care. Since May 2018, the type 2 opt-out has been replaced by the ND opt-out, and the type 1 is not communicated as an option for the citizens anymore. The type 1 decisions will be removed in 2020.

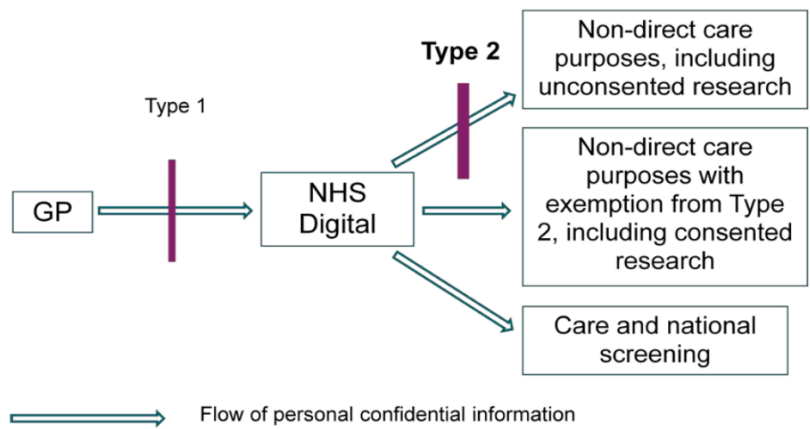

**Figure 1**: The "Old" Opt-Out System in England.[50]

---

[47] See Streams. Available at: https://deepmind.com/applied/deepmind-health/working-partners/how-were-helping-today/. Accessed 10 June 2019.
[48] McGoogan (2017).
[49] NHS (National Health Service) Digital is an executive non-departmental public body of the Department of Health in the UK. The NHS Digital is the national provider of information, data and IT systems for commissioners, analysts and clinicians in health and social care. Available at: https://www.gov.uk/government/organisations/nhs-digital/about. Accessed 10 June 2019.
[50] NHS Digital: Implement Type 2 patient opt-outs (2016), p. 2.

As Figure 1 above shows, the type 1 opt-out is the only option, which truly stops data from being shared outside direct care. The type 2 option had several unclear limitations. It stopped data from being shared outside of NHS Digital for research and planning. The citizens might have expected from the opt-out information that their confidential information can be circulated and used only for care and planning inside the NHS. However, pseudonymized data was an exemption, since it could be further processed for a secondary purpose regardless of the patients' choice. Probably the reason for NHS to push citizens toward the type 2 opt-outs was the need for data for planning and facilitating research. The goal was to reduce the costs of care and improve the effectiveness of the healthcare system, thus *nudging* the patients for using the type 2 opt-out had a public interest.

### 4.1.2 The New National Data Opt-Out System

From May 2018, the type 2 opt-out has been replaced by the ND opt-out. The previously recorded type 2 opt-outs have been automatically converted to ND opt-outs. Existing type 1 opt-outs will be respected until 2020, when the Department of Health will remove them. What is more important, the government stopped promoting type-1 opt-outs for citizens. There are leaflets, posters and an information webpage about the new ND opt-out, and neither of them conveys the fact that the patients are still able to choose the type 1 option at the GPs, which would truly stop the processing of their confidential information outside of direct care. The ND opt-out is communicated for patients in several ways such as: by healthcare staff, leaflets, posters, online, and they could also get information via telephone. The type 1 opt-out is not publicized to citizens on these publicly available materials. Furthermore, on the information website, which is on these materials, the type 1 opt-out is not even mentioned. Information about the type 1 opt-out[51] can be only found on the NHS website, under the "resources for health and care staff," by clicking on the "more information for patients with a previous type 2 opt-out" menu.[52] This communication indicates that the government intends to slowly roll out the type 1 option, before the final removal in 2020.

---

[51] NHS Digital: Opting out of sharing your confidential patient information. Available at: https://digital.nhs.uk/about-nhs-digital/our-work/keeping-patient-data-safe/how-we-look-after-your-health-and-care-information/your-information-choices/opting-out-of-sharing-your-confidential-patient-information. Accessed 15 June 2019.
"Type 1 opt-out: medical records held at your GP practice: You can also tell your GP practice if you do not want your confidential patient information held in your GP medical record to be used for purposes other than your individual care. This is commonly called a type 1 opt-out. This opt-out request can only be recorded by your GP practice."
[52] NHS Digital, Opting out of sharing your confidential patient information. Available at:
https://digital.nhs.uk/about-nhs-digital/our-work/keeping-patient-data-safe/how-we-look-after-your-health-and-care-information/your-information-choices/opting-out-of-sharing-your-confidential-patient-information. Accessed 10 June 2019.

The new ND opt-out will only apply when identifiable data is shared for research and managing the efficient and safe operation of the healthcare system,[53] thus in the case of anonymized and pseudonymized data, this choice does not apply. Another issue is that the use of the terms "anonymized," "pseudonymized" and "de-identified" is not consistent in the UK legal terminology. These terms are used interchangeably in various codes, white papers and leaflets. Therefore, the citizens might be confused about their meaning, and they might assume that anonymization means their data is safe. However, in the UK, the term "anonymization" has the same meaning as "de-identification" in the General Data Protection Regulation (GDPR), thus it can also involve "pseudonymization." In other words, "anonymized" data might be just "pseudonymized" data in many cases. Pseudonymization is the separation of data from the direct identifiers (e.g., name, address, NHS number), so that re-identification is not possible without additional information (the "key") which is held separately. Thus, it is still possible to re-identify the data subjects after this security measure.[54]

In this chapter, we use the terms "anonymization"[55] and "pseudonymization"[56] in a manner consistent with the GDPR and the EU Data Protection Board. According to these sources, anonymized data can no longer lead to the identification of the data subject, while pseudonymization is a useful security measure which reduces the "linkability" of a dataset with the original data subject.[57] The umbrella term for both of these measures is "de-identification." Another possible ground of misunderstanding for the citizens is the notion of "direct care." They might assume this activity consists of their care by the professionals in a healthcare institution. However, direct care is a much broader concept,[58] since it may also include the assurance of safe and high-quality care, which requires a background work that might not be apparent for the patients.

### *4.2 The Presentation of the Information on the Information Materials for Patients*

---

[53] NHS Digital, About the national data opt-out. Available at: https://digital.nhs.uk/services/national-data-opt-out-programme. Accessed 10 June 2019.
[54] Article 29 Working Party, Opinion 05/2014 on Anonymisation Techniques (2014), p. 3.
[55] Regulation (EU) 2016/679 of the European Parliament of 27 April 2016 on the protection of natural persons with regard to the processing of personal data and on the free movement of such data, and repealing Directive 95/46/EC (General Data Protection Regulation) [hereinafter "GDPR"]. Recital 26: anonymous information, namely information which does not relate to an identified or identifiable natural person or to personal data rendered anonymous in such a manner that the data subject is not or no longer identifiable.
[56] Article 4 (5) of the GDPR: "pseudonymisation" means the processing of personal data in such a manner that the personal data can no longer be attributed to a specific data subject without the use of additional information, provided that such additional information is kept separately and is subject to technical and organizational measures to ensure that the personal data are not attributed to an identified or identifiable natural person.
[57] Article 29 Working Party, "Opinion 05/2014 on Anonymization Techniques" (WP216, 10 April 2014), p. 20.
[58] National Data Guardian (2013), p. 128. "A clinical, social or public health activity concerned with the prevention, investigation and treatment of illness and the alleviation of suffering of individuals. It includes supporting individuals' ability to function and improve their participation in life and society. It includes the assurance of safe and high-quality care and treatment through local audit, the management of untoward or adverse incidents, person satisfaction including measurement of outcomes undertaken by one or more registered and regulated health or social care professionals and their team with whom the individual has a legitimate relationship for their care."

Electronic copy available at: https://ssrn.com/abstract=4087791

**4.2.1 General Information**

The UK government needs the citizens' health data for research and policy planning; thus it is crucial how the information about opt-out is presented to them. As explained above, the ND opt-out is communicated to patients in several ways: by healthcare staff, leaflets, posters, videos, online, and they may also obtain information via telephone. These materials contain mostly similar information. However, there are special versions of them for young people and minorities. In this section, we introduce how these materials communicate the secondary use of health data for citizens to help them to make their decision. Since the default setting is the sharing of health data, these materials do not have to convince citizens to give their consent, just to understand the importance of their data and accept the situation.

On all the information materials, the first thing which is presented to patients is the value of their health information: "Information about your health and care helps us to improve your individual care, speed up diagnosis, plan your local services and research new treatments."[59] By starting with this information, the patients might realize the importance and public interest behind the secondary use of their data, thus there might be less of a chance they choose to opt-out. There is also a possibility that after reading this information, they might not continue to read over the whole of the information material.

The NHS materials about the secondary use of confidential patient data continue with the following statement: "In May 2018, the strict rules about how this data can and cannot be used were strengthened." This statement is debatable since the exemptions of the opt-out system did not change: the de-identified data can be still used for a new purpose even in the case of opt-out, and the type-1 opt-out is planned to be canceled. The NHS also promises on the leaflet that: "The NHS is committed to keeping patient information safe and always being clear about how it is used." However, the Google Deepmind case highlighted that the NHS might also share confidential patient data with private companies for "direct care" which turned out not to be a proper legal basis[60] after the investigation by the Information Commissioner.[61]

The information materials state that "You can choose whether your confidential patient information is used for research and planning." This statement is true, however, only for the type 1 opt-out, which is about to be cancelled, and not communicated to patients. By this presentation, the patients might be biased and assume the ND opt-out can provide the full protection of their data. The leaflet explains the meaning of confidential patient information as follows: "information identifies you and says something about your health, care or treatment. Information that only identifies you, like your name and address, is not considered confidential patient information and may still be used: for example, to contact you if your GP practice is merging with another."[62]

---

[59] NHS (2018) Your Data Matters to the NHS, p. 1. Available at: https://digital.nhs.uk/services/national-data-opt-out-programme/supporting-patients-information-and-resources. Accessed 25 May 2019.
[60] Powles and Hodson (2017), pp. 351-367.
[61] The ICO has ruled the Royal Free NHS Foundation Trust failed to comply with the Data Protection Act when it provided patient details to Google DeepMind. Available at: https://ico.org.uk/about-the-ico/news-and-events/news-and-blogs/2017/07/royal-free-google-deepmind-trial-failed-to-comply-with-data-protection-law/. Accessed 10 June 2019.

As it was highlighted in the previous section, the misunderstandings around de-identified information may lead to misinterpretation. The leaflet clearly explains how information can be confidential, and the processing of it is necessary for administrative purposes and direct care, even in the case of opt-out. In the next part, the leaflet explains who can use this confidential patient information: "NHS, local authorities, universities and hospital researchers, medical colleges and pharmaceutical companies researching new treatments."[63] What this part of the information leaflet does not specify, are other not medical-related companies, such as Google Deepmind and other IT corporations that might also get access to health data. As healthcare is becoming digitized, such as X-ray diagnostics using AI and machine learning, IT companies are gaining a crucial role in providing the backbone of direct care and medical research. For many years, medical doctors played the most important role in improving healthcare. However, very recently programmers and analytics have contributed significantly to improve the quality of care.[64]

### 4.2.2 Information About the Exemptions

After introducing the potential users of health data, the information leaflet turns to situations, when the opt-outs might be ignored. It only introduces one situation with the highest public interest, the epidemics: "You can choose to opt out of sharing your confidential patient information for research and planning. There may still be times when your confidential patient information is used: for example, during an epidemic where there might be a risk to you or to other people's health." However, there are many other situations when the opt-outs might be ignored, such as court orders and with regard to the use of de-identified data. The information about other exemptions can be only found on the NHS opt-out website.[65]

The information materials clarify that confidential patient information will be used for direct care regardless of the opt-outs: "Will choosing this opt-out affect your care and treatment? No, your confidential patient information will still be used for your individual care. Choosing to opt-out will not affect your care and treatment. You will still be invited for screening services, such as screenings for bowel cancer."[66] On the one hand, this information empowers citizens to opt-out if they wish, since it does not affect their individual care. On the other hand, this information may also inform patients that their information will be used regardless of their decision.

---

[62] NHS (2018) Your Data Matters to the NHS, p. 2. Available at: https://digital.nhs.uk/services/national-data-opt-out-programme/supporting-patients-information-and-resources. Accessed 25 May 2019.
[63] NHS (2018) Your Data Matters to the NHS, p. 2. Available at: https://digital.nhs.uk/services/national-data-opt-out-programme/supporting-patients-information-and-resources. Accessed 25 May 2019.
[64] Meskó, Hetényi and Győrffy (2018), pp. 1-4.
[65] NHS, When your choice does not apply. Available at: https://www.nhs.uk/your-nhs-data-matters/where-your-choice-does-not-apply/. Accessed 10 June 2019.
[66] NHS (2018) Your Data Matters to the NHS, p. 2. Available at: https://digital.nhs.uk/services/national-data-opt-out-programme/supporting-patients-information-and-resources. Accessed 10 June 2019.

The following information might be the main nudging strategy against opt-out: "What should you do next? You do not need to do anything if you are happy about how your confidential patient information is used. If you do not want your confidential patient information to be used for research and planning, you can choose to opt-out securely online or through a telephone service."[67] Most citizens do not have deep knowledge on how their information is processed. The time and effort[68] to read through privacy policies are burdensome[69] and in many cases time wasting, since there might be no alternative choices (e.g., MS Windows, national health services). Thus, people have their subjective perception of the quality of the health service they directly receive, but less understanding concerning the information processing and administrative work behind it. It is possible that the patient is satisfied with the service, but her data is misused by private companies; and on the contrary, it is also possible that the patient had a bad experience during treatment, and chose not to share data, which would have been wisely used by private corporations to improve her future care.

### 4.2.3 Special Opt-Out Materials

The NHS is providing special information materials for minorities and young people. The information leaflets are slightly different from the general materials. The most apparent difference is the design of the information documents. While the general leaflet has a blue background with white letters, the others which target minorities and young people look more like traditional materials, with black and blue letters on white background. The materials do not have a significant difference in length: they are approximately 450-500 words, stretched across two pages. The leaflets for minorities and young people have cover pages, making the material more comfortable to hold and read. However, the wording of the materials is slightly different, since the leaflet for young people has the shortest general word length, which makes it easier to understand.

|  | General | Minority information leaflet | Young people |
| --- | --- | --- | --- |
| number of pages | 2 | 4 | 4 |
| number of words | 449 | 504 | 505 |
| number of characters (without space) | 2317 | 2548 | 2275 |
| number of characters (with spaces) | 2694 | 3033 | 2755 |
| number of spaces | 377 | 485 | 480 |
| **general word length** | **5.16** | **5.05** | **4.50** |

---

[67] NHS (2018) *Your Data Matters to the NHS*, p. 2. Available at: https://digital.nhs.uk/services/national-data-opt-out-programme/supporting-patients-information-and-resources. Accessed 10 June 2019.
[68] McDonald and Cranor (2008), pp. 543-568.
[69] Solove (2013), pp. 1880-1903.

**Table 1:** Comparison of NHS Information Materials on the ND Opt-Out System.

There is special information material for care givers (e.g., doctors, nurses),[70] which advises them to promote the NHS website providing information for patients on the value of health data. The NHS website addresses the concerns about privacy and the selling of data to private companies: "While people may feel uncomfortable with the idea of the NHS 'selling' data, there would also be concerns if valuable data is given away to companies for free."[71] The other part of the leaflet with instructions to the healthcare staff is neutral, asking them to disclose both the risks and advantages for patients and to support them in their choice.[72] Overall, the special materials contain similar information, but they communicate the same message in a slightly different way.

## 5 Opt-Out Statistics

The NHS regularly releases statistics on opt-outs.[73] From May 2018, the type 2 opt-out has been replaced by the ND opt-out, and the existing type 2 opt-outs have been automatically converted to ND opt-outs. In March 2018, the number of type 2 opt-outs was 1.4 million representing 2.42 % of the population. At the time of writing this chapter, the latest publication of the opt-out statistics[74] was in March 2019, and the total number of ND opt-outs was 1.6 million, which represented 2.74 % of the population of England. This number is relatively low, especially if the well-advertised ND opt-out system is compared with the previous system.

As explained before, the type-1 opt-out is no longer advertised to the public. Furthermore, there are no statistics released on the type 1 opt-outs in 2019, and the existing type 1 opt-outs are planned to be removed in 2020. In March 2018, the number of type 1 opt-outs was 1.85 million, which was 3.13 percent of the population in England.

---

[70] NHS (2018) A guide to the national data opt-out for carers. Available at: https://digital.nhs.uk/services/national-data-opt-out-programme/supporting-patients-information-and-resources. Accessed 11 May 2019.
[71] Understanding Patient Data, Frequently Asked Questions. Available at: https://understandingpatientdata.org.uk/what-you-need-know. Accessed 23 June 2019.
[72] NHS (2018) A guide to the national data opt-out for carers, p. 1. "If you think the person you care for is happy for their information to be shared you don't need to do anything further." "If you think that the person you care for would wish to register a national data opt- out, or you are satisfied that registering a national data opt-out is in that person's best interest then follow step 3."
[73] NHS Digital, statistics on the volumes of national data opt-outs. Available at: https://digital.nhs.uk/data-and-information/publications/statistical/national-data-opt-out. Accessed 12 June 2019.
[74] NHS Digital, statistics on the volumes of national data opt-outs, March 2019. Available at: https://digital.nhs.uk/data-and-information/publications/statistical/national-data-opt-out/march-2019/ndop-mar19. Accessed 12 June 2019.

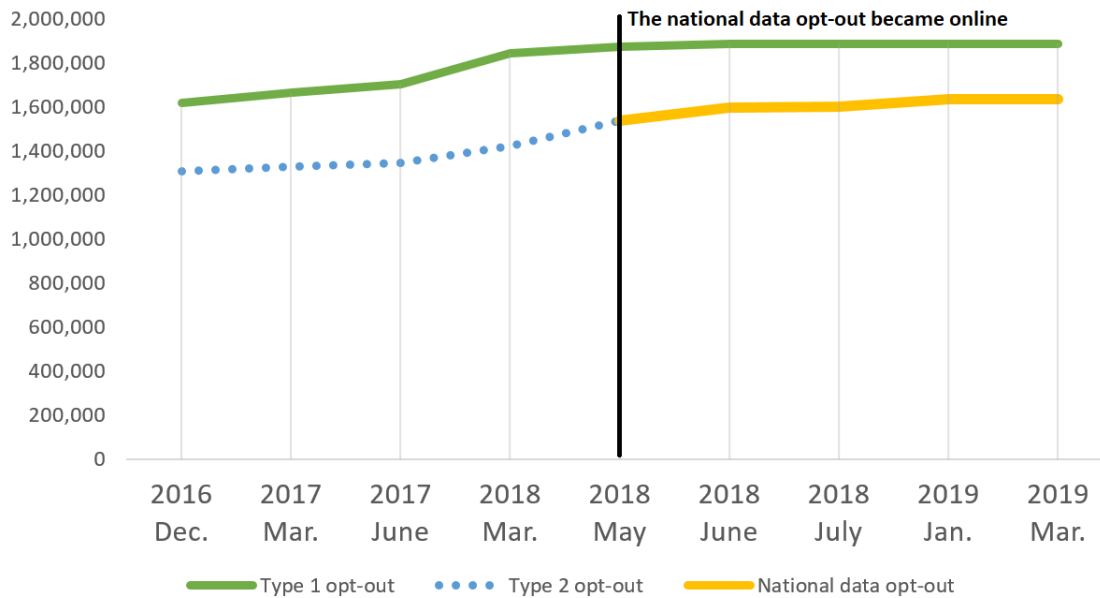

**Figure 2**: Opt-Outs Rates in England.

The type 1 and type 2 opts-outs were presented at the GP level before May 2018.[75] The number of type 2 opt-outs is accurate since every opt-out choice has been registered and reported from the GPs to NHS Digital with identifiable patient information; thus, there are no duplications. However, the number of type 1 opt-outs is not accurate, since one person may have been recorded several times at different GPs, and the GPs could only report the number of opt-outs to NHS, without personal data.[76] After informing the public about the ND opt-outs in many different ways (e.g., leaflets, posters, videos), the number of opt-outs has not raised significantly, since only 0.3 % of the population decided not to share their personal data in the new system. The slight increase might indicate that the NHS's campaign on the importance of sharing health data was successful, and their message reached the patients.

|  | **Number of opt-outs** | **Rate (compared to the population of England)** |
|---|---|---|
| Type 1 opt-outs (March 2018) | 1,846,250 | 3.13 % |

---

[75] Type 1 opt-outs have been reported as instances (i.e., number of times the opt-out code occurs within GP records, which may include the same patient recorded at more than one practice), therefore the NHS Digital could not de-duplicate this information.

[76] NHS Digital, statistics on the volumes of national data opt-outs, March 2018. Available at: https://digital.nhs.uk/data-and-information/publications/statistical/care-information-choices/mi-care-information-choices-england-march-2018. Accessed 16 May 2019.

| | | |
|---|---|---|
| Type 2 opt-outs (March 2018) | 1,422,250 | 2.42 % |
| **May 2018** | ND opt-out is online, and the existing type-2 opt-outs are converted to ND opt-outs | |
| ND opt-outs (July 2018) | 1,602,910 | 2.71 % |
| ND opt-outs (March 2019) | 1,639,012 | 2.74 % |

**Table 2:** Opt-Out Rates in England.

# 6 Hard Paternalism in Healthcare?

The gist of the matter is whether the UK government may legitimately nudge its citizens and whether this could be seen as a kind of "hard paternalism." In *Why Nudge? The Politics of Libertarian* Paternalism, Cass Sunstein focuses on finding a justification for various nudging techniques. On the normative level, Sunstein tries to challenge John Stuart Mill's "harm principle,"[77] which suggests that individuals can do whatever they want as long as their actions do not harm others. If they do, the government intervention can be justified to constrains the person and prevent such harm.[78]

The harm principle is justified on the grounds that individuals know better what is good for them and that governments do not have enough information and resources to know all the needs of its citizens. Sunstein refers to this argument as the "epistemic argument" and indicates that this way of reasoning is sometimes wrong. Sunstein goes on and argues that in some cases, paternalistic interventions are desirable, specially where people are likely to make a mistake and it is necessary to provide the means for improved decision-making.[79]

Paternalism comes in many forms. Sunstein explains that there are mainly four types of paternalism grouped into two main categories. One category differentiates "hard" versus "soft" paternalism. One of the criteria for distinguishing these types of paternalism relates to the imposition of material costs on individuals. "Hard paternalism" occurs when people's free choice is coerced by the government, whereas "soft paternalism" is when the person is free to choose the form of action. The latter denotes very little or no intervention from the state[80] such as in the GPS example explained above in Sect. 2.

In traditional "hard paternalism," the so-called "nanny state" uses its coercive power to nudge its citizens to do what is in their best interests.[81] Nonetheless, "soft paternalism" holds the view that government intervention is legitimate and justified only when the person is consciously aware and acts voluntarily.[82] Mill's famous example of the person who is about to cross a damaged bridge (so-called "Bridge Exception"), illustrates this case in point clearly.[83] Consider the case where the government could not communicate the risks of a bridge that is about to collapse because of language limitations (i.e., the intervened person does not speak the local language and thus she can neither read the signs nor understand any warning signals given). In this scenario, the government's use of force to stop the person from crossing the bridge would be justified as her liberty consists in doing what she wants, and falling and dying is most probably not her will in this case.[84]

---

[77] Mill (1859).
[78] Corrales and Jurčys (2016), p. 534.
[79] Corrales and Jurčys (2016), p. 534.
[80] Corrales and Jurčys (2016), p. 534.
[81] Bishop (2009), p. 296.

Sunstein provides some compelling arguments against those who oppose government intervention, nudging, and autonomy. He criticizes "welfarist" objections for failing to take into account the fact that most public policy decisions are already made and individuals have only a limited ability to control the exercise of those underlying choices. Moreover, "welfarists" fail to acknowledge the empirical findings on behavioral economics as well as the fact that choice architectures and nudges are everywhere. Sunstein advocates for a choice architecture, which he labels *libertarian paternalism*: individuals are encouraged to make active choices, which helps to solve the shortcomings of a much criticized, one-size-fits-all approach.[85]

# 7 Conclusion

Behavioral Law and Economics has become increasingly relevant as a point of reference in policy-making and regulation over the past decade. In this regard, Nudge Theory lends itself as powerful tool that can help us to improve the normative framework. At first glance, there seems to be no major differences in the new ND-opt-out system. However, a closer look reveals some subtle, but very important changes. The bone of contention is that choices provided in the new ND-opt-out are going to be more limited in 2020, as the type-1 opt-out will disappear. Moreover, the UK's government attempt to nudge patients by framing (or even hiding) information seems to fall under the "hard paternalism" category in the Sunstein narrative, which is not the most desirable one.

The linchpin and value of nudges is to bring to the fore the prospect of "choice." Therefore, the best way for the UK government to act paternalistic, but at the same time respecting citizen freedom is with a "soft paternalism" approach. This approach does not seem to run counter to the use of active choice mechanisms, nudges or default rules. On the contrary, the paternalist actions may lead to welfare gains that are greater than the welfare loss. Patients may feel frustrated to discover that certain decisions have already been made for them. Therefore, it is important to leave an option to revisit those decisions if people do not like them. This is the reason why we advocate to keep the type-1 opt-out system as a default rule. In this way, the opt-out choice architecture will enable patients with more options. This, in turn, will offer a stronger data protection scheme without overriding individual freedom.[86]

---

[82] Tanner (2007), p. 200; Hartley (2012), p. 70; Angner (2016), p. 264.
[83] See, e.g., generally, Jackson (2006), pp. 68-69.
[84] Sunstein (2014), pp. 63-99.
[85] Corrales and Jurčys (2016), pp. 534-535.

**Acknowledgement:** This research is supported by a Novo Nordisk Foundation grant for a scientifically independent Collaborative Research Program in Biomedical Innovation Law (grant agreement number NNF17SA027784) and the Multidisciplinary Health Cloud Research Program: Technology Development and the Application of Big Health Data. Academia Sinica, Taipei, Taiwan.

---

[86] Corrales and Jurčys (2016), pp. 534-535.